# Heterojunction Vertical Band-to-Band Tunneling Transistors for Steep Subthreshold Swing and High ON Current

Kartik Ganapathi[a] and Sayeef Salahuddin[b]

Department of Electrical Engineering and Computer Sciences,

University of California Berkeley, CA – 94720, USA.

## ABSTRACT

**We propose a Heterojunction Vertical Tunneling FET and show using self-consistent ballistic quantum transport simulations that it can provide very steep subthreshold swings and high ON current, thereby improving the scalability of Tunnel FETs for high performance. The turn-on in pocket region of the device is dictated by modulation of heterojunction barrier height. The steepness of turn-on is increased because of simultaneous onset of tunneling in the pocket and the region underneath and also due to contribution to current by vertical tunneling in the pocket. These factors can be engineered by tuning heterojunction band offsets.**

a) kartik@eecs.berkeley.edu

b) sayeef@eecs.berkeley.edu

I. INTRODUCTION

The interest in Band-to-Band Tunneling Field Effect Transistors (TFETs) has gained traction over the recent years with their demonstration in various material systems of the ability to provide Subthreshold Swings (SS) steeper than 60 mV/decade at room temperature – a fundamental limit in classical MOS devices - thereby providing a possible route to scaling down voltage and power [1]-[4]. However, the presence of tunneling resistance in the ON state of the device severely inhibits the scalability of TFETs for high performance. One of the proposed device structures to overcome this limitation involves having a heavily doped halo (pocket) region in the gate-to-source overlap region of a conventional p-i-n TFET geometry and hence introducing an additional component of tunneling current due to band overlap along the body thickness, amounting to an increased tunneling area [5]-[8]. From hereon for brevity, we will refer to this device structure as Vertical TFET (VTFET) indicating that the dominant tunneling current component is in the direction normal to the semiconductor-dielectric interface, while the regular p-i-n TFET structure will be referred to as Lateral TFET (LTFET) denoting the fact that the tunneling is predominantly in the direction of transport. Recent electronic transport simulations of III-V VTFETs have not shown significant improvement in turn-on characteristics in comparison to LTFETs although an increased ON current is observed [7]. In this letter, we propose a VTFET with wide band gap material in the channel region (Hetero-VTFET from now on), in contrast to the Heterojunction LTFET [9]-[12], and show using self-consistent ballistic quantum transport simulations with realistic multi-band Hamiltonian that such a device can offer greatly improved OFF state and turn-on behavior while still providing high ON current owing to pocket geometry. We investigate the physics underlying the turn-on characteristics and the factors governing steepness of SS. Our results provide insight into directions for optimization of VTFET geometry thereby improving TFET scalability.

II. SIMULATION APPROACH

The structure of the simulated device is shown in Fig. 1(a). The doping concentrations in source, pocket and drain are 5 x $10^{19}$, 5 x $10^{19}$ and 5 x $10^{18}$ cm$^{-3}$ respectively. The asymmetry in doping concentrations of source and drain is motivated by the lower conduction band density of states in III-V materials and the need to inhibit ambipolar conduction. The channel is assumed to have an n-type doping of $1 \times 10^{15}$ cm$^{-3}$ to account for unintentional doping arising due to defects. The source, channel and drain are 20 nm long each while the pocket is 10 nm wide and 3.6 nm deep. We use a body thickness of 10 nm. A 1.2 nm thick dielectric with $\kappa$ = 15.4 is used as the gate insulator. The crystallographic direction for transport is (100) and the body is confined along (001) direction. For comparison between device structures, the workfunction difference between semiconductor and gate is adjusted so that the OFF state occurs at same gate voltage. The alloy composition of In$_x$Ga$_{1-x}$As is chosen such that the band offsets of the channel with that of the source facilitate simultaneous onset of lateral and vertical tunneling in the device – details of which will be explained in a subsequent section.

We use a 4 x 4 k. p Kane's second order Hamiltonian to describe the band structure of InAs and In$_{0.53}$Ga$_{0.47}$As [13]. The spurious states in dispersion relations – an artifact arising due to confinement - are removed following the prescription in [14]. Spin-orbit coupling is neglected to reduce computation time. Also, we ignore the effect of strain in our heterostructures. The band gaps of InAs and In$_{0.53}$Ga$_{0.47}$As are thus estimated to be 0.5437 and 0.9130 eV respectively. Band-offsets are assumed to be in the same ratio as that of the corresponding bulk materials.

We perform 2-D ballistic transport simulations within the Non Equilibrium Green's Function (NEGF) formalism with the said multiband Hamiltonian description, thereby implicitly accounting for all tunneling paths. In all our simulations, both gates are maintained at identical electrostatic potential. The width of the device is assumed to be large. Hence a periodic boundary condition is used and the momentum modes are summed numerically in calculation of charge densities and current. The electrostatic potential, electron and hole concentrations are computed self-consistently by iteratively solving Poisson and NEGF equations.

III. RESULTS AND DISCUSSION

A. Transfer Characteristics

Fig. 1(b) shows the plot of drain current $I_D$ as a function of gate voltage $V_G$ for homogenous VTFET (Homo-VTFET henceforth) and Hetero-VTFET devices. Similar characteristics are shown for Homo and Hetero LTFETs on the same plot for reference purposes. A drain voltage $V_D$ of 0.4 V is used throughout. Evident from the characteristics is the fact that Hetero-VTFETs provide significant reduction in OFF state leakage as compared to their homogenous counterparts. They also provide much steeper subthreshold swings (a minimum SS of 16 mV/decade vs. 62 mV/decade in Homo-VTFETs). Although a reduction in ON current as expected is observed due to staggered bands at the heterojunction, the pocket geometry, by virtue of boosting the current due to vertical tunneling, reduces this penalty in comparison to LTFETs.

*B. OFF State*

The OFF state behavior of a Hetero-VTFET can be understood by examining the energy band diagram and energy resolved current $I(E)$ $(= T(E) \times (f_1(E)-f_2(E)))$ where $T(E)$ is the transmission coefficient at a given energy $E$ and $f_1$ and $f_2$ the Fermi Dirac distributions corresponding to source and drain electrochemical potentials respectively) plotted in Figs. 2(a) and (b) respectively for Homo and Hetero-VTFETs. The smaller OFF state leakage in Hetero-VTFETs is due to two reasons – i) larger barrier height due to band discontinuities and ii) larger effective mass $m^*$ in the channel ($0.046m_0$ at the bottom of conduction band in $In_{0.53}Ga_{0.47}As$ vs. $0.032m_0$ in InAs, $m_0$ being the free electron mass) implying a smaller wavefunction penetration.

*C. Turn-on mechanism*

We now turn to explain a fundamental difference in the turn-on mechanism of a VTFET in comparison to that of an LTFET. The potential profiles along the length of the device in Hetero-VTFET near the semiconductor - top gate insulator interface are plotted in Fig. 2(c) for three different gate voltages in the region where the turn-on is the steepest. We note that this steep turn-on corresponds to the transition wherein the peak in $I(E)$, shown for corresponding gate voltages in Fig. 2(d), moves from below the barrier at the heterojunction to above it. It can be seen from the inset of Fig. 2(c) that the tunneling width in the energy region where most of the current flows, remains virtually unchanged during this transition. This is in stark contrast to the case of LTFET turn-on where barrier *width* and not *height* is modulated. Two important clarifications are in order here – i) The steep turn-on is *not* sensitive to the sharpness of the barrier at the heterojunction interface and any broadening of this barrier introduced by a scattering mechanism will manifest only as a shift in the turn-on voltage. This is due to the fact that the $p^+$-$n^+$ tunnel junction at the source-pocket interface is always turned on, by virtue of doping, as can be seen in Fig. 2(c). ii) While the turn-on mechanism should be identical in Homo-VTFET, the phenomenon is masked because of a large lateral tunneling current akin to LTFET current that flows underneath the pocket as is evidenced by the presence of tunneling states in a plot of logarithmic Local Density of States (LDOS) along the length of Homo-VTFET, near the semiconductor-back gate insulator interface, shown in Fig. 3(b).

### D. Factors Affecting Steepness

There are two other reasons for greater steepness in turn-on of a Hetero-VTFET – i) the onset of lateral tunneling underneath the pocket together with tunneling in the pocket region. Fig. 3(a) shows logarithmic LDOS plots in Hetero-VTFET near the back-gate for $V_G$ = 0.55 V and 0.6 V. We observe that owing to the optimum choice of band discontinuities between source and channel materials, unlike Homo-VTFET where the onset of lateral tunneling precedes turn-on in pocket region as explained before, in case of Hetero-VTFET it occurs simultaneously. ii) The contribution of vertical tunneling component of current during steep turn-on. It must be observed that while vertical tunneling contributes to a larger ON state current in case of Homo-VTFETs too, it does not lead to steeper swing as it sets in at large gate-voltages wherein the device is already turned-on. However, since turn-on voltage in Hetero-VTFETs is larger owing to heterojunction band-offsets, vertical tunneling does contribute to steepness of SS. The LDOS plots along the body thickness in the middle of pocket region of Homo and Hetero-VTFETs during their steepest turn-on, shown in Figs. 3(c) and 3(d) respectively indicate the presence of greater density of tunneling states in the latter.

### E. Negative Transconductance

An interesting feature of the $I_D$-$V_G$ characteristics of Hetero-VTFET is a region of negative transconductance between $V_G$ = 0.775 V and 0.85 V. Our simulations show that the negative transconductance arises due to resonance between tunneling states in the pocket region and states in the channel – the density of which is non-monotonic near the bottom of conduction band, due to difference in dispersion relations of InAs and $In_{0.53}Ga_{0.47}As$. The higher current at lower $V_G$ is due to higher density of states in the channel coupling to tunneling states in the pocket that carry majority of the current. The subsequent increase in $I_D$ for large $V_G$ is due to the appearance of increasing number of tunneling states in the pocket, many of which propagate through states far above the bottom of $In_{0.53}Ga_{0.47}As$ conduction band in the channel.

IV. CONCLUSION

Our simulations show that the turn-on in a Hetero-VTFET is governed by the modulation of heterojunction barrier height as opposed to tunneling width modulation in case of LTFETs. We also demonstrate that by choosing the right band offsets, the steepness of turn-on can be increased owing to - i) simultaneous onset of tunneling in the pocket and region underneath it and ii) vertical tunneling contribution. The negative transconductance region that can occur in Hetero-VTFET characteristics is explained to be due to resonant tunneling between states in the pocket and the channel. Although in this study we have restricted ourselves to nominal device structures so as to retain our emphasis on underlying physics, we note that enhanced performance can be obtained by optimizing the critical device parameters like heterojunction band offsets (tuned using Ga mole fraction), pocket width and depth and doping profiles in source and pocket regions, thus fully leveraging the advantages of Hetero-VTFETs.

ACKNOWLEDGMENTS

This work was supported in part by FCRP center on Functional Engineered and Nano Architectonics (FENA) and National Science Foundation (NSF). Computing resources was provided by NSF Center for Computational Nanotechnology (NCN).

FIGURE CAPTIONS

Figure 1. (a) Schematic of the simulated Homo and Hetero-VTFET devices. (b) $I_D$-$V_G$ characteristics of Homo and Hetero-VTFETs at $V_D$ = 0.4 V. Similar characteristics are shown for the case of Hetero LTFET for comparison sake. The Hetero LTFET has top-gate extending over the body alone. The steepest SS obtained in case of Homo and Hetero-VTFETs are 62 mV/decade (from $V_G$ = 0.375 V to 0.4 V) and 16 mV/decade (from $V_G$ = 0.575 V to 0.6 V) respectively. The black and green curves in (b) are identical to blue and black curves of Fig. 2(a) in [7] respectively except for a shift in OFF state voltage done for comparison purposes.

Figure 2. (a) Energy band diagrams in Homo and Hetero-VTFETs at z = 0 (semiconductor- top gate insulator interface) at $V_G$ = 0.25 V which corresponds to OFF state. (b) The corresponding energy resolved current *I(E)*. (c) Energy band diagrams in Hetero-VTFET at z = 0 for three gate voltages in the region of steep turn on. The inset has the same zoomed in the pocket-channel interface region. The arrow in the inset shows the energy window where majority of tunneling current flows. (d) *I(E)* in Hetero-VTFET for same gate voltages as in (c).

Figure 3. (a) Plots of logarithmic LDOS in Hetero-VTFET at z = 10 nm (semiconductor- back gate insulator interface) at $V_G$ = 0.55 V and 0.6 V. (b) LDOS plot in Homo-VTFET at z = 10 nm at $V_G$ = 0.375 V. (c) and (d) LDOS plots along body thickness at x = 15 nm (middle of pocket region) in Homo ($V_G$ = 0.375 V) and Hetero ($V_G$ = 0.575 V) VTFETs respectively.

FIGURES

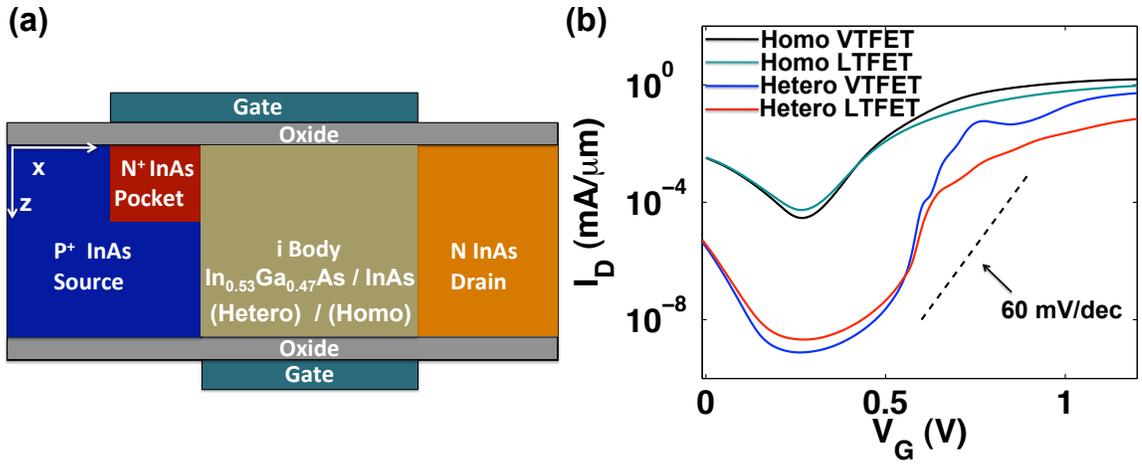

Figure 1

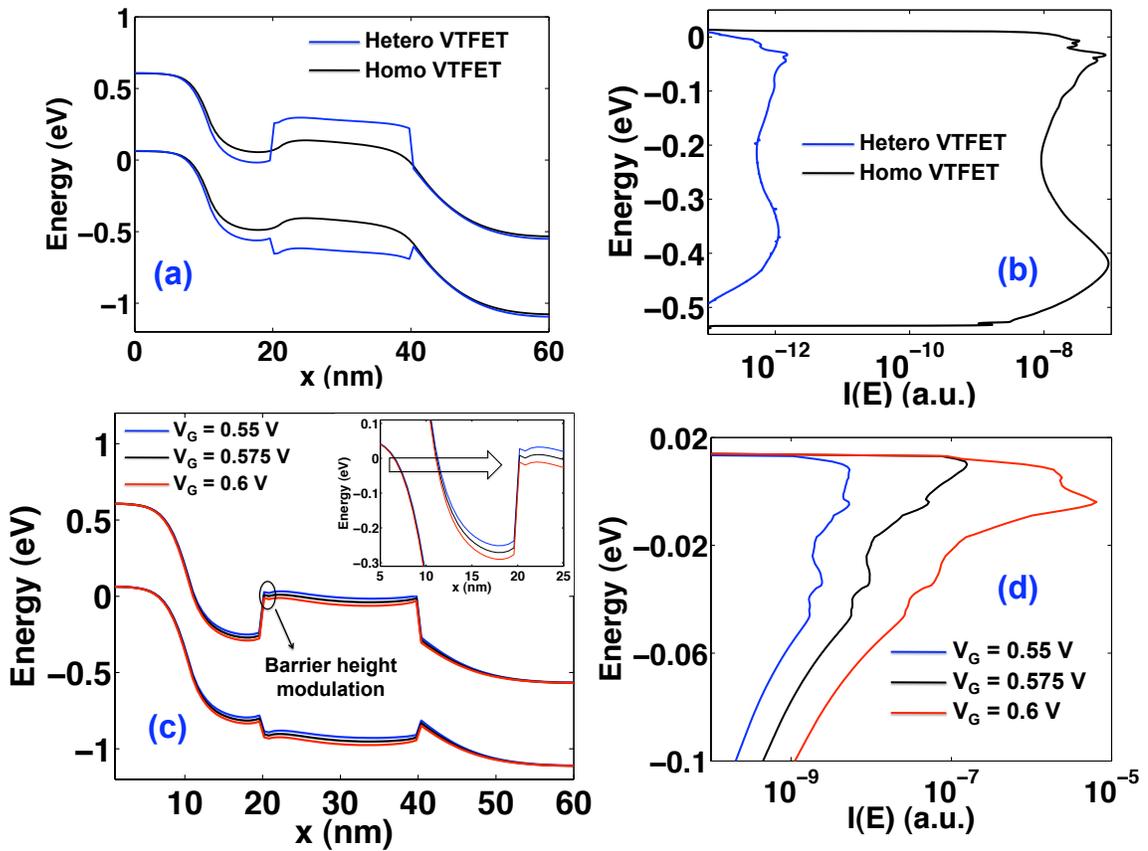

Figure 2

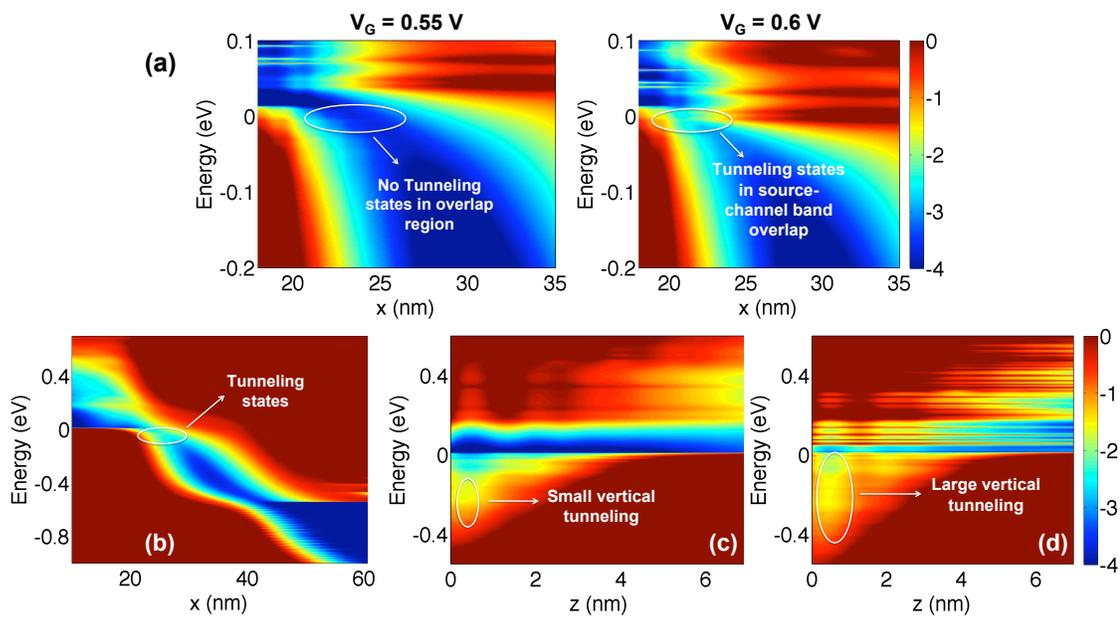

Figure 3